\documentclass[aps, jcp, prb,  two column,showpacs,groupedaddress]{revtex4-1}
\usepackage{float}
\usepackage{graphicx} 
\usepackage{subfigure}
\usepackage{amsmath}
\usepackage{dcolumn}
\usepackage{multirow}
\usepackage{bm}     
\usepackage{amssymb,amssymb}   
\usepackage[left= 1 in, right= 1 in, top= 1 in , bottom= 1 in]{geometry}
\usepackage{hyperref}
\usepackage{xcolor} 
\hypersetup{ colorlinks, citecolor=blue ,linkcolor=black }

\usepackage{chemfig}

\usepackage{hyperref}

\hypersetup{linktocpage}

\begin{document}
	\vspace*{0.35in}

	\title{Molecular Dynamics Study of  Transport Properties of Cysteine in Water}
	
	\author{Hem Prasad Bhusal}
	
	\author{ Narayan Prasad Adhikari }
	\email{npadhikari@tucdp.edu.np}
	\affiliation{Central Department of Physics, Tribhuvan University, Kirtipur, Kathmandu, Nepal}
	
	\begin{abstract}
		\noindent Molecular dynamics simulation is a prominent way of analyzing the dynamic properties of a system. The molecular dynamics simulation of diffusion, an important transport property, of dilute solution of cysteine in SPC/E water at five different temperatures (288 K, 293 K, 303 K, 313 K, 323 K) under the pressure of 1 bar is studied using GROMACS. OPLS-AA force field parameters are used throughout the simulation. The system under study consists of 3 cysteine molecules (mole fraction 0.003) as solute and 1039 water molecules (mole fraction 0.997) as solvent. The radial distribution functions (RDF's) for five different combinations of atoms of solvent-solvent and solute-solvent molecules are studied for structural analysis. At least two or more distinct peaks are observed in RDF's plots implying that there are interactions between atoms of solvent-solvent and solute-solvent at least up to two co-ordination shells. The self-diffusion coefficients of solute and solvent are determined exploiting mean square displacement (MSD) in Einstein's equation. The self-diffusion coefficients are used to calculate the binary diffusion coefficients by means of Darken's relation. The calculated values of self-diffusion and binary diffusion coefficients are compared with available experimental values and they agreed within 12\% error. The temperature dependency of diffusions are demonstrated via Arrhenius plots and with the help of these plots activation energies for diffusions are calculated which agreed with experimental results within 13\% error. 
		
	\end{abstract}
	
	\maketitle
	
	\section*{Introduction}
	\noindent Biomolecules are the organic molecules present in the living organisms, including large macromolecules: proteins, carbohydrates, lipids and nucleic acids, and small molecules such as metabolites~\cite{wikibio}. Amino acids, important biomolecules~\cite{wikiamino}, are the organic compounds containing amine ($\--$NH$_2)$ and carboxyl ($\--$COOH) as
	functional groups as well as hydrogen (H) or alkyl (R) as side chain.  They are structural units of protein. Cysteine which is abbreviated as Cys or C is a proteinogenic sulfur-containing non-essential amino acid~\cite{wikicys}. Its linear formula is HOOCCH(NH$_2$)CH$_2$SH, which contains an alpha amino group, a carboxyl group, and a side chain consisting of a mercaptomethyl group. The IUPAC name of cysteine is 2-amino-3-mercaptopropanoic acid.
	
\noindent Cysteine is a white crystalline solid having molar mass 121.15 gram per mole and melting point is 513 K~\cite{wikicys}. Its solubility in water is 16 gram per 100 mL at 288 K~\cite{lcystop}. But cysteine exhibits hydrophobic nature, due to which it generally resides in the interior of proteins~\cite{biocys}. Cysteine is a non-essential amino acid as it is synthesized in the human body under the ordinary physiological state if there is adequate amount of methionine, which is also a sulfur-containing amino acid. By virtue of ability to form disulfide bonds (methionine can't form this bond), cysteine plays a crucial role in protein structure and in protein-folding mechanisms~\cite{wikicys}. Also cysteine is essential for the synthesis of highly anti-oxidative  Glutathione which is important in the detoxification and protection of various tissues and organs in the body~\cite{aminolcys}. Further, cysteine supports in the absorption of nutrients from intestine, metabolism of lipids, enhancing the fertility, strengthening the immune system, preventing from the diseases like dementia, multiple sclerosis as well as Parkinson~\cite{aminolcys}, etc. It is also recognized as anti-aging amino acid. All these activities put the cysteine in special position that can not be substituted by any other amino acid.

\noindent The term transport phenomena means the process by which the mass, linear momentum, angular momentum, energy and charge are transferred from one part of the system to another due to non-uniformity  or inhomogeneity in the system. Diffusion, an important transport property, is the phenomenon of transfer of mass as a result of random molecular motion. Various experimental techniques, viz., Peak-Height method~\cite{wenrui},  Nuclear Magnetic Resonance (NMR)~\cite{virk}, etc. as well as Molecular Dynamics (MD) simulations~\cite{virk} have been performed to study the diffusion phenomenon of amino acids in water. These studies were mainly concerned about the effect of concentration, polarity, temperature on the diffusion of amino acids. However, to the best of our knowledge, cysteine diffusion in water using MD simulation has not been performed yet.\\

\section*{Diffusion}
\noindent Diffusion is a process by which matter is transported from one part of system to another part as a result of random molecular motion
due to concentration gradient or thermal agitation without any external force and bulk
motion~\cite{J.crank}. Diffusion plays many important roles in non-living substances as well as living organisms. The diffusion in a homogeneous system having no chemical concentration gradient is called self-diffusion and the corresponding diffusion coefficient is termed as self-diffusion coefficient~\cite{H.hira}. In order to calculate the self-diffusion coefficient one of the most common method is exploiting the Einstein's equation which relates diffusion coefficient with mean square displacement (MSD) of the particles as;
\begin{equation}
\label{selfdiff}
D = \lim_{t \to \infty}  \frac{\langle \lbrack r(t) - r(0) \rbrack ^2 \rangle}{6 t}.
\end{equation}
In equation (\ref{selfdiff}), $r(t) - r(0)$ is the displacement of particle from reference point during the course of time $t$, $ \lbrack r(t) - r(0) \rbrack ^2$ is the square of displacement and $\langle ... \rangle$ represents the ensemble average and hence ${\langle \lbrack r(t) - r(0) \rbrack ^2 \rangle}$ gives MSD of particle.\\

\noindent Binary diffusion is the diffusion of two different substances in their binary mixture and the resulting diffusion coefficient is called binary diffusion coefficient. The Darken's relation which is used to calculate the binary diffusion coefficient is given by
\cite{lsdarken},
\begin{equation}
\label{darkeneq}
D_{12}=N_2D_1+N_1D_2. 
\end{equation}
In equation (\ref{darkeneq}), $D_{12}$ is the binary diffusion coefficient, $D_1$ and $D_2$ are the self-diffusion coefficients of substances 1 and 2 respectively, and $N_1$ and $N_2$ are the corresponding mole fractions.

 \section*{Computational Details}
 \subsection*{Modeling of System}
\noindent The one of the best simulation technique to study the dynamic properties of substance at molecular level is molecular dynamics. Here we provide the initial positions and velocities of each particles; the new positions of the particles at each time step can be obtained using Leapfrog algorithm in the simulation. Hence, molecular simulation can provide the trajectories of particles which are utilized to find the equilibrium and transport properties of substances~\cite{smith}. For $N$ interacting particles in the classical MD simulation Newton's equation of motion given in equation (\ref{newtoneq}) is solved to find the new position of the particles.
\begin{equation}
	\label{newtoneq}
m_i\frac{\partial^2\textbf{r}_i}{\partial t^2} = -\nabla_i U(r) = \textbf{F}_i.
\end{equation}
In equation (\ref{newtoneq}), $m_i$ is the mass and $r_i$ is the position of $i^{th}$ particle. In RHS of this equation, the force on $i^{th}$ particle is equal to negative gradient of the potential.

\noindent We start the simulation work through modeling of system where we assign the molecules in terms of mass, charge, bond, angle, dihedral, etc. All these essential parameters are specified in selected force field, which is OPLS-AA in our case. The atoms in the system is treated as spherically symmetric potentials in classical force fields. The force between each particle is calculated from pairwise additive potential functions~\cite{allen}. The total potential energy $U_{total}$ of the system is due to contributions from bonded interactions and non-bonded interactions between individual atoms/molecules of the system. The bonded interactions are classified as bond stretching, bond angles, dihedral angle interactions. Lennard-Jones interactions and Coulomb interactions come under the category of non-bonded interactions~\cite{gromanual}. Hence, the total potential energy function in our system is written as,
\begin{equation}
\label{totalpotential}
U_{total} = U_{bond} + U_{angle} + U_{proper} + U_{LJ} + U_{Coulomb}.
\end{equation}

\subsection*{Simulation Set Up} 

\noindent We have simulated the 3 cysteine and 1039 water molecules at five different temperatures. Cysteine is  three carbon containing alpha-amino acid. There are 14 atoms in a single molecule of cysteine. Different atoms possesses different partial charge due to difference in electro-negativity. Also the same atom possesses different partial charge based on the group of attachment. The Coulomb interaction occurs due to the partial charge exist in atoms/molecules. Likewise, the Vander Waal's interaction occurs as a result of induced dipole interaction. The parameters for non-bonded interactions: partial charge, sigma and epsilon as well as for bonded interactions: bond lengths, angles and dihedrals, are assigned in the OPLS-AA force field by default. In case of water, the required force field parameters are specified in water model. We have used SPC/E water model in our simulation. In this model each hydrogen atom of water is specified with a partial charge of +0.4238 e, and the oxygen atom is assigned with the partial charge of $\--$0.8476 e, where e is the electronic charge whose magnitude is $1.6022 \times 10^{-19}$ Coulomb. All the parameters for SPC/E water model are included in the file \textit{spce.itp} inherent to GROMACS.

\begin{table} [H]
	\centering
	\label{forcefieldforH2O}
	\caption{Force field parameters for SPC/E water model.}
	\begin{tabular}{|c|c|} \hline
		Parameters &Values\\
		\hline
		$K_{OH}$ &  3.45 \ $\mathrm{x}$ \ 10$^5$ \ kJ mol$^{-1}$ nm$^{-2}$ \\
		$b_{OH} $ & 0.1 nm \\
		$K_{HOH}$ & 3.83 \  $\mathrm{x}$ \ 10$^2$  \ kJ mol$^{-1}$ rad$^{-2}$\\
		$\Theta_o $ & $ 109.47^0$ \\
		\hline
	\end{tabular}
\end{table}

\noindent The non-bonded parameters are listed in the file \textit{nonbonded.itp} which are presented below:

\resizebox{0.47\textwidth}{!}{\begin{tabular}{ccccccc}
			[\textbf{atomtypes}]&& &&&&\\

 atoms& at.num &   mass   &     charge &  ptype & sigma     & epsilon \\
 
 N & 7 & 14.0067 & $\--$0.90 & A & 3.3000e-01 & 7.1128e-01\\
 CA & 6 & 12.0110 & 0.12 & A & 3.5000e-01 & 2.7614e-01\\
 C& 6 & 12.0110 & 0.52 & A& 3.75000e-01   &  4.3932e-01\\
O & 8& 15.9994& $\--$0.53& A  & 3.0000e-01 & 7.1128e-01 \\
CB  & 6  & 12.0110& 0.06 & A& 3.5000e-01 & 2.7614e-01\\
SG & 16&32.0600 &$\--$0.45& A  &3.5500e-01 &1.0460e+00\\
 H1  & 1 &1.0080 & 0.36 & A  &0.0000e+00 &0.0000e+00 \\
 H2  & 1 &1.0080 & 0.36 & A  &0.0000e+00 &0.0000e+00 \\
HA&1&1.0080 & 0.06&A&2.5000e-01  & 1.2552e-01 \\
 HO  & 1 &1.0080 & 0.45 & A  &0.0000e+00 &0.0000e+00 \\
 HB1&1&1.0080 & 0.06&A&2.5000e-01  & 1.25520e-01\\
HB2&1&1.0080 & 0.06&A&2.5000e-01  & 1.2552e-01\\
 HG  & 1 &1.0080 & 0.27 & A  &0.0000e+00 &0.0000e+00 \\
 OC & 8& 15.9994& $\--$0.44& A  & 2.9600e-01 & 8.7864e-01 \\
OW&8&15.9994&$\--$0.82 & A &3.1656e-01 & 6.5019e-01 \\
 HW1&1&1.0080 &0.41   & A  &0.0000e+00& 0.00000e+00\\
HW2&1&1.0080 &0.41   & A  &0.0000e+00& 0.0000e+00\\
 
\end{tabular}}\\

\noindent In this table, first column represents name of atoms. Second, third and fourth columns respectively give the atomic number, atomic mass (in atomic mass unit) and partial charge (in e) of respective atoms. Sixth column specifies particle type, where A stands for atom. Seventh and eighth column give the sigma and epsilon parameters of corresponding atom respectively. For dissimilar atoms pair, the values of sigma and epsilon are calculated by using following relations in OPLS force field \cite{gromanual}.
\begin{equation}
\sigma_{ij} = (\sigma_{ii} \sigma_{jj})^{1/2}, \quad
\epsilon_{ij} = (\epsilon_{ii} \epsilon_{jj})^{1/2}.
\label{ljconversion}
\end{equation}
Once after enclosing three cysteine molecule in cubic simulation box, we add 1039 water molecules in the box to form cysteine-water system. We employ periodic boundary condition which eliminates the surface effect. After solvation of cysteine in water, system is subjected to the process of energy minimization; which brought the system to one of a local minima where system possesses minimum energy. In GROMACS, 
we used the steepest-descent algorithm for energy minimization.
 \begin{figure}[H]
 	\centering
 \includegraphics[scale=0.48]{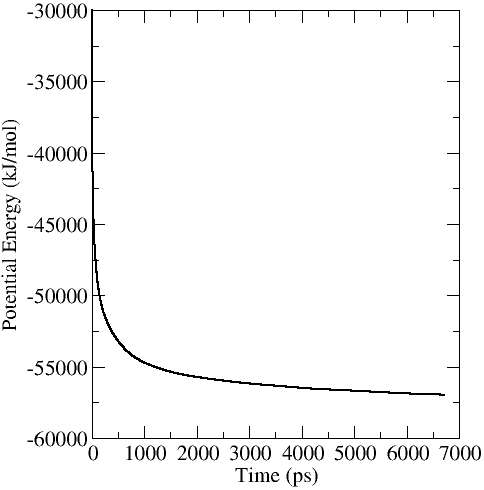}
 	\caption{Potential energy after energy minimization.}
 	\label{potentialenergy}
 \end{figure}
 
 \begin{figure}[H]
 	\centering
 \includegraphics[scale=0.35]{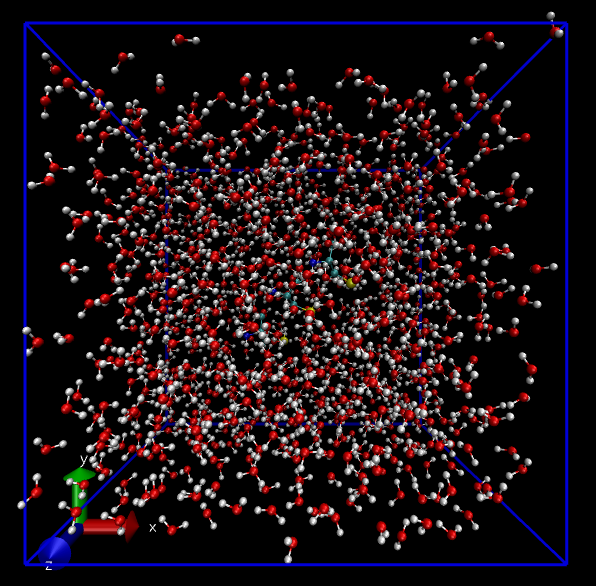}
 	\caption{System after energy minimization.}
 	\label{fig:emafter}
 \end{figure}
\noindent Figure (\ref{potentialenergy}) is the energy minimization curve for our system. The energy of the system is said to be minimized if the maximum force experienced in the system is less than \textit{emtol}, which is 50 kJ mol$^{-1}$ nm$^{-1}$. Also for a stable system total energy of the system should be negative. In our system we got that:\\
\textbf{Potential energy} = $\--$5.6946953 $\times$ 10$^{4}$ kJ mol$^{-1}$  and steepest descent converged to $ {F_{max}}$ $ < 50$ kJ mol$^{-1}$ nm$^{-1}$ after 6713 steps.

\noindent Since the total potential energy is negative and nearly constant, our system is stable. Figure (\ref{fig:emafter}) shows our system after energy minimization.

\noindent After energy minimization, the system is ready to study the dynamic properties. However, the dynamical properties such as diffusion, viscosity usually vary with the parameters like temperature, pressure, density, etc. Therefore, before starting the production run (where the physical properties will be calculated) these fore-mentioned parameters should be kept constant during simulation and the system under study has to brought in the state of thermodynamic equilibrium. The process of bringing the system in thermodynamic equilibrium state is called equilibration or npt run. The temperature in the system is kept constant (thermal equilibrium) by using thermostat and the pressure is kept constant by using barostat. In GROMACS, the process of adjusting temperature and pressure to the desired values are respectively termed as \textit{temperature coupling} and \textit{pressure coupling}.

\noindent In the equilibration parameter file i.e. \textbf{equilibration.mdp}, we have used \textit{md integrator}, which is leapfrog algorithm, with step size of 0.001 ps for 5 $\times$ 10$^7$ steps in all runs. This results equilibration time 50 ns. We have used PME (Particle-mesh Ewald) for long-range Coulomb interaction with Fourier spacing of 0.12 nm and cut-off distance of 1.0 nm. V-rescaling have been used for \textit{temperature coupling} with time constant of 0.01 ps. Run is carried out at five different temperatures: 288 K, 293 K, 303 K, 313 K and 323 K. \textit{Pressure coupling} have been done through Barendsen barostat to reference pressure of 1 bar. The \textit{pressure coupling} time constant and isothermal compressibility have been respectively taken as 0.8 ps and 4.6 $\times$ 10$^{-5}$ bar$^{-1}$. We have applied the LINCS constraint algorithm to convert all bonds into constrains during equilibration run. LINCS algorithm is default in GROMACS~\cite{gromanual} if there is not specified any algorithm explicitly.

\begin{table}[H]
	\caption{Temperatures and densities of system after equilibration run.}
	\label{table:npt temp and dens}
\resizebox{0.475\textwidth}{!} {\begin{tabular}{|c|c|c|c|c|}
		\hline
Coupling        & Equilibrium    & Equilibrium & Estimated \\
Temperature (K) & Temperature (K) & Density (kg m$^{-3}$) & Density (kg m$^{-3}$)~\cite{waterdensity} \\
		\hline
288 & 287.992 $\pm$ 0.005 & 1001.120 $\pm$ 0.036 & 999.102 \\
		\hline
293 & 292.988 $\pm$ 0.005 & 998.586 $\pm$ 0.041 & 998.207 \\
		\hline
303 & 302.992 $\pm$ 0.008 & 993.397 $\pm$ 0.031 & 995.650 \\
		\hline
313 & 312.994 $\pm$ 0.006 & 987.487 $\pm$ 0.061 & 992.219 \\
		\hline
323 & 322.988 $\pm$ 0.012 & 980.920 $\pm$ 0.057 & 988.039 \\
		\hline
		
\end{tabular}}
\end{table}

\noindent Table (\ref{table:npt temp and dens}) listed out the equilibration temperatures and densities of the system at different reference or coupling temperatures. The equilibrium temperature is almost equal to reference temperature. As our system is very dilute we have compared density of system with that of water at different temperatures. We have found that the densities of system are consistent with that of water within error less than 1\%.

\noindent After finishing the equilibration run, the system is ready for production run or the simulation of NVT ensemble. In this phase of simulation, we determined the diffusion coefficient. Basically \textbf{nvt.mdp} file is similar to the parameter file of equilibration run. However, in nvt run we did not apply the \textit{pressure coupling} since we are executing NVT ensemble. Further it is not required to generate the initial velocities in nvt run as the simulation continues with the velocities generated in equilibration run.

\section*{Results and Discussion}
\noindent In this section, we present the energy profile of system obtained after the production run, structural analysis and transport property-diffusion coefficients of constituents of the system at different temperatures.

\section*{Energy Profile} 
\noindent Different interactions exist in the system result various types of energies. The manifestation of all the energies present in the system caused by different interactions constitute the energy profile of the system. The interactions associated with bond stretches, angle vibration, dihedral rotations are categorized as bonded interactions and the energies corresponding to these interactions are termed as bonded interaction energies. The energy terms associated with non-bonded interactions are Coulomb potentials and LJ potentials. Both bonded and non-bonded interaction energies contribute to the total potential energy of the system and on addition of total potential energy and kinetic energy finally yield total energy of the system at particular temperature.

\begin{figure}[H]
	\centering
	\includegraphics[scale=0.35]{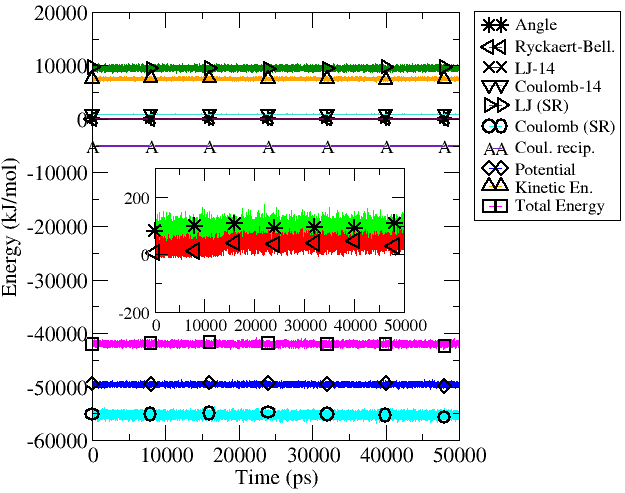}
	\caption{Energy profile of system at 288 K.}
	\label{enprof288}
\end{figure}
\noindent Figure (\ref{enprof288}) shows the energy profile of the system at 288 K. This energy profile portrays that the Coulomb and LJ interactions are the prominent interactions contributing to the total potential energy. The energy values due to bonded interactions (angle vibrations and R-B dihedrals) and due to pair interactions (LJ-14 and Coulomb-14) are very little. These values almost coincide to zero with respect to other energy term values in the graph. Therefore, the contribution to total potential energy from these interactions is negligible. The magnitude of Coulomb potential energy is very large with negative sign. This drives the total potential energy to be negative. Similarly, large negative value of potential energy and small positive value of kinetic energy  lead the system to acquire negative total energy after production run. This implies our system is bound and stable. Also the relationship between kinetic energy and temperature of the system is shown via table (\ref{table:ke and t}).

\begin{table}[H]
	\caption{Relationship between kinetic energy and temperature.}
	\label{table:ke and t}
	\resizebox {0.48 \textwidth }{!} {\begin{tabular}{|c|c|c|c|}
		\hline
		SN        & Temperature (T)   & Kinetic Energy (K.E.)  & $\frac{K.E.}{T}$ \\
		& (K) &  (kJ mol$^{-1}$) & (kJ mol$^{-1}$ K$^{-1}$)\\
		\hline
		1 & 288 & 7564.40 & 26.265 \\
		\hline
		2 & 293 & 7695.65 & 26.265 \\
		\hline
		3 & 303 & 7958.39 & 26.265 \\
		\hline
		4 & 313 & 8221.05 & 26.265 \\
		\hline
		5 & 323 & 8483.59 & 26.265 \\
		\hline
		
	\end{tabular}}
\end{table}

\noindent In the table (\ref{table:ke and t}), we have calculated the ratios of kinetic energies obtained after production run and the corresponding absolute temperatures. All these ratios have given the same value 26.265 kJ mol$^{-1}$ K$^{-1}$. That is,

$\frac{K.E.}{T}$ = a constant\\

\noindent Or, K.E.  = a constant $\times$ T\\
\noindent This implies kinetic energy is proportional to the absolute temperature. 

\section*{Structure of System}
\noindent The radial distribution functions, RDF's, of pair of atoms are used for the study of structure of system. From the analysis of distribution of neighboring atoms around a reference atom, structural properties of the system can be determined. The effect of temperature in the structure of system is also studied by analyzing RDF's at five different temperatures.

\subsection*{Structural Analysis of Solvent}
\noindent We have used normal water as solvent. Since the LJ-parameters of hydrogen atoms of water are zero, therefore, we study the RDF of oxygen atom of water for the structural analysis. Here we have discussed the RDF of two oxygen atoms of water molecules at five different temperatures. The RDF plots depicted that how the atoms are organized from reference atom.

\begin{figure}[H]
	\centering
	\includegraphics[scale=0.33]{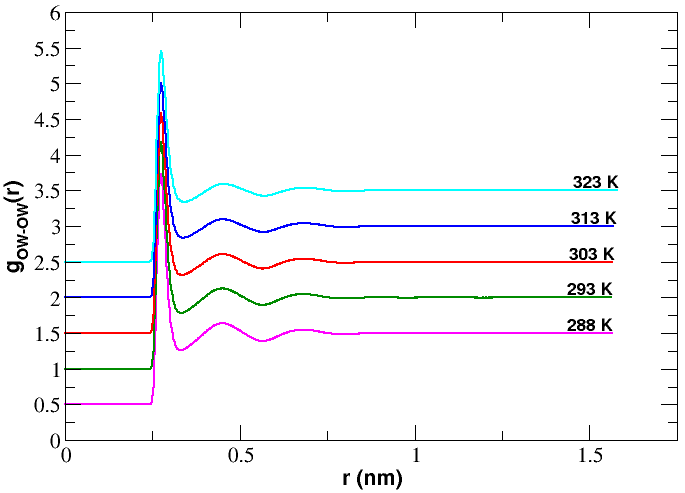}
	\caption{RDF plot of oxygen-oxygen of water molecules, $g_{OW\--OW}$(r), at different
		temperatures.}
	\label{rdfow}
\end{figure}

\begin{figure}[H]
	\centering
	\includegraphics[scale=0.37]{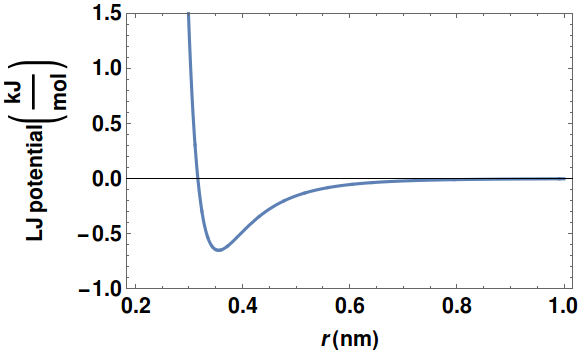}
	\caption{LJ potential plot as a function of distance for OW$\--$OW of water.}
	\label{ljowow}
\end{figure}

\begin{table}[H]
	\centering
	\caption{Details of RDF of oxygen-oxygen atoms of water at different temperatures.}
	\label{tablerdfow}
	\resizebox {0.475 \textwidth }{!}{%
		\begin{tabular}{|c|c|c|c|c|c|c|c|}\hline
			\multicolumn{8}{|c|}{RDF analysis of OW$\--$OW atoms }\\    \hline
			T (K) & ER (nm) & FPP (nm)   &     FPV & SPP (nm) & SPV     & TPP (nm) & TPV \\ \hline
			288 & 0.240 & 0.274 & 3.230 & 0.450 & 1.140 & 0.682 & 1.049\\ 
			\hline
			293 & 0.240 & 0.274 & 3.175 & 0.450 & 1.129 & 0.680 & 1.048\\ 
			\hline
			303 & 0.240 & 0.274 & 3.077 & 0.450 & 1.110 & 0.686 & 1.043\\ 
			\hline
			313 & 0.240 & 0.276 & 3.000 & 0.450 & 1.096 & 0.690 & 1.041\\ 
			\hline
			323 & 0.240 & 0.276 & 2.945 & 0.450 & 1.091 & 0.686 & 1.037\\ 
			\hline
		\end{tabular}}
	\end{table}
\noindent In the figure (\ref{rdfow}), there are three distinct peaks. The first peak, which is located at the separation of about 0.27 nm from centered atom's position, is highest and sharpest. This implies that at this position maximum number of oxygen atoms are clustered from the reference oxygen atom. In other words, the probability of finding oxygen atoms at the first peak position is highest. This is the most preferable position or minimum energy position from the centered atom. From the figure (\ref{ljowow}) of LJ potential of OW$\--$OW of water, we fond the $\sigma$ value of water oxygen is 0.316557 nm and the Vander Waal's radius is 2$^\frac{1}{6} \sigma \approx$ 0.36 nm. However, the FPP in our case is 0.27 nm less than 0.36 nm. This reveals the fact that there is not only LJ interaction between oxygen atoms of water but also other interactions such as Coulomb and bonded interactions are present.

\noindent The second and third peaks are relatively shorter and wider, which are located approximately at positions 0.45 nm and 0.68 nm respectively. Excluded region, in which RDF is zero, has extended up to 0.24 nm from the center of reference oxygen atom. Any other oxygen atom can not exist within excluded region due to strong repulsive forces: namely $r^{-12}$ term of LJ interaction and repulsive Coulomb interactions~\cite{ishworpaper}. We have also studied the effect of temperature on RDF. Figure (\ref{rdfow}) shows that with increase in temperature the peak positions are shifted to right, heights of peaks are decreased and widths are increased. This reflects that our system have became less organized with increase in temperature. The increase in thermal agitation of atoms in the system with rising temperature accounts this fact. Further beyond third peak graph is straight line possessing unit value on average. This indicates there is no pair correlation of oxygen atoms.

\subsection*{RDF of OC and OW}
\noindent Here OC refers to carbonyl oxygen atom of cysteine and OW means oxygen atom of water. The RDF of OC and OW, $g_{OC\--OW}(r)$, gives the insight about how the carbonyl oxygen atoms of cysteine organized around the oxygen atom of water.

\begin{figure}[H]
	\centering
	\includegraphics[scale=0.33]{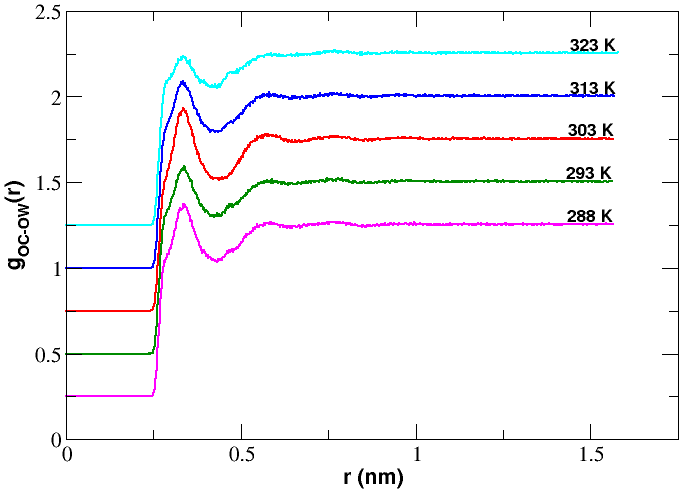}
	\caption{RDF plot of carbonyl oxygen of cysteine and oxygen of water, $g_{OC\--OW}$(r), at different
		temperatures.}
	\label{rdfoc1}
\end{figure}

\begin{table}[H]
	\centering
	\caption{Details of RDF of carbonyl oxygen of cysteine and oxygen of water at different temperatures.}
	\label{tablerdfoc}
	\resizebox {0.475 \textwidth }{!}{%
		\begin{tabular}{|c|c|c|c|c|c|c|c|}\hline
			\multicolumn{8}{|c|}{RDF analysis of OC$\--$OW atoms }\\    \hline
			T (K) & ER (nm) & FPP (nm)   &     FPV & SPP (nm) & SPV     & TPP (nm) & TPV \\ \hline
			288 & 0.240 & 0.336 & 1.125 & 0.584 & 1.018 & 0.746 & 1.020\\ 
			\hline
			293 & 0.242 & 0.338 & 1.094 & 0.590 & 1.020 & 0.778 & 1.024\\ 
			\hline
			303 & 0.242 & 0.336 & 1.185 & 0.578 & 1.031 & 0.762 & 1.021\\ 
			\hline
			313 & 0.242 & 0.336 & 1.092 & 0.580 & 1.026 & 0.768 & 1.022\\ 
			\hline
			323 & 0.242 & 0.336 & 0.985 & 0.586 & 1.003 & 0.766 & 1.020\\ 
			\hline
		\end{tabular}}
	\end{table}
	
	\noindent Figure (\ref{rdfoc1}) has two distinct peaks (first and second) and third peak is not so sharp. The first peak, which is located at the separation of about 0.33 nm from the position of reference oxygen atom of water, is highest and sharpest. This implies that at this position maximum number of carbonyl oxygen atoms of cysteine clustered from the reference oxygen atom. Therefore, this is the most preferred position of carbonyl oxygen atoms to cluster around the oxygen atom of water. The second and third peaks are relatively shorter and wider, which are located approximately at the positions 0.58 nm and 0.76 nm respectively. Excluded region extends up to 0.24 nm from the center of reference oxygen atom. There is not possible to find any carbonyl oxygen within excluded region due to strong repulsive forces. Figure (\ref{rdfoc1}) shows with increase in temperature the peak positions are shifted to right, heights of peaks are decreased and widths are increased. This reflects that the less organization of carbonyl oxygen atoms around the reference oxygen atom with increase in temperature. Further beyond third peak graph is straight line possessing unit value on average. This indicates there is no pair correlation of carbonyl oxygen atoms and the reference oxygen atom of water.

\subsection*{RDF of HO and OW}
\noindent Here HO refers to hydroxyl hydrogen atom of carboxyl group of cysteine and OW means oxygen atom of water. The RDF of HO and OW, $g_{HO\--OW}(r)$, gives the insight about how the hydroxyl hydrogen atoms of cysteine organized around the oxygen atom of water.

\begin{figure}[H]
	\centering
	\includegraphics[scale=0.33]{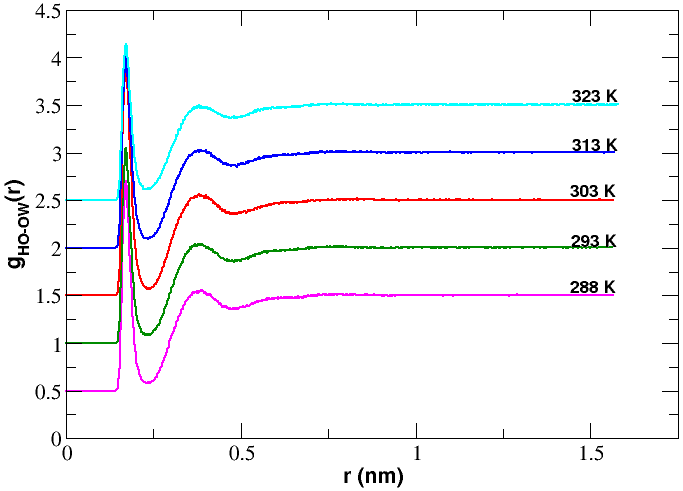}
	\caption{RDF plot of hydroxyl hydrogen of cysteine and oxygen of water, $g_{HO\--OW}$(r), at different
		temperatures.}
	\label{rdfho}
\end{figure}

\begin{table}[H]
	\centering
	\caption{Details of RDF of hydroxyl hydrogen of cysteine and oxygen of water at different temperatures.}
	\label{tablerdfho}
	\resizebox {0.475 \textwidth }{!} {%
\begin{tabular}{|c|c|c|c|c|c|}
	\hline
\multicolumn{6}{|c|}{RDF analysis of HO$\--$OW atoms }\\    \hline
T (K) & ER (nm) & FPP (nm)   &     FPV & SPP (nm) & SPV \\ 
\hline
			288 & 0.142 & 0.172 & 2.198 & 0.386 & 1.055 \\ 
			\hline
			293 & 0.140 & 0.172 & 2.043 & 0.378 & 1.043 \\ 
			\hline
			303 & 0.140 & 0.172 & 2.468 & 0.382 & 1.061 \\ 
			\hline
			313 & 0.140 & 0.172 & 2.055 & 0.380 & 1.030 \\ 
			\hline
			323 & 0.140 & 0.172 & 1.640 & 0.380 & 0.995\\ 
			\hline
		\end{tabular}}
	\end{table}
	
	\noindent Figure (\ref{rdfho}) has two distinct peaks. The first peak, which is located at the separation of about 0.17 nm from the position of reference oxygen atom of water, is higher and sharper. This implies that at this position maximum number of hydroxyl hydrogen atoms of cysteine clustered from the reference oxygen atom. Therefore, this is the most preferred position of hydroxyl hydrogen atoms to cluster around the oxygen atom of water. The second peak is shorter and wider than first one, which is located approximately at the position 0.38 nm. Excluded region extends up to 0.14 nm from the center of reference oxygen atom. There is not possible to find any hydroxyl hydrogen within excluded region due to strong repulsive forces. Figure (\ref{rdfho}) shows with increase in temperature the peak positions are shifted to right, heights of peaks are decreased and widths are increased. This reflects that the less organization of hydroxyl hydrogen atoms around the reference oxygen atom with increase in temperature. Further beyond second peak graph is straight line possessing unit value on average. This indicates there is no pair correlation of hydroxyl hydrogen atoms and the reference water oxygen atom.

\section*{Diffusion Coefficients}
	
\subsection*{Self-Diffusion Coefficient of Cysteine}
\noindent The self-diffusion coefficients of cysteine have been calculated for five different temperatures by using corresponding MSD curves. We have determined the self-diffusion coefficient by dividing the slope of linear nature of MSD plot by six according to Einstein's equation (\ref{selfdiff}). In order to get the linear nature of MSD curve, we have plotted the MSD curves for 3 ns for all temperatures even though the production run was done for 50 ns.

\begin{figure}[H]
	\centering
	\includegraphics[scale=0.32]{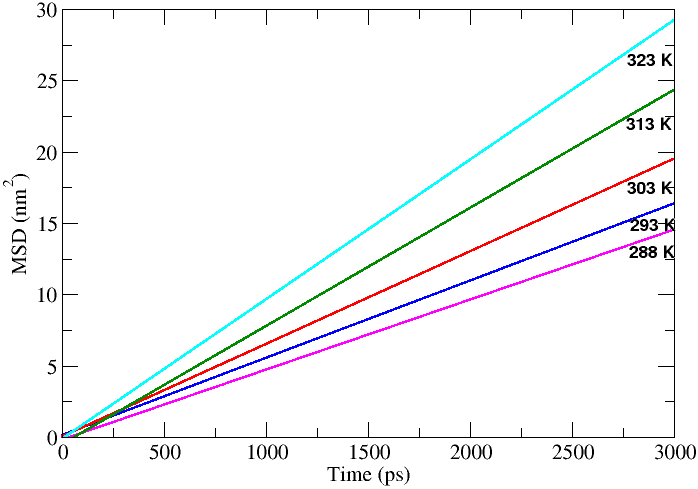}
	\caption{MSD vs time plot of cysteine at different temperatures.}
	\label{msdunkcomb}
\end{figure}

\noindent Figure (\ref{msdunkcomb}) shows the MSD versus time plot at different temperatures. From this figure, it is seen that as the temperature increased slope of the MSD curves have also been increased which in turn increased the self-diffusion coefficient. This is reflected in the following table.

\begin{table}[H]
	\caption{Simulated values of self-diffusion coefficient of cysteine at different temperatures.}
	\label{table:cys_self_diff}
\resizebox {0.48 \textwidth } {!} { \begin{tabular}{|c|c|}
		\hline
		Temperature (K) & Self-Diffusion Coefficient $D_C^{sim}$ ( $10^{-10}$ m$^2$/s)\\
		\hline
		288 & 8.17 $\pm$ 1.10 \\
		\hline
		293 & 9.01 $\pm$ 0.27\\
		\hline
		303 & 10.81 $\pm$ 0.41 \\
		\hline
		313 & 13.78 $\pm$ 2.79 \\
		\hline
		323 & 16.29 $\pm$ 0.51\\
		\hline
	\end{tabular}}
\end{table}

\subsection*{Self-Diffusion Coefficient of Water}
\noindent As previously for cysteine, we have studied MSD plots and hence calculated self-diffusion coefficient of water at different temperatures using the same approach.

\begin{figure}[H]
	\centering
	\includegraphics[scale=0.32]{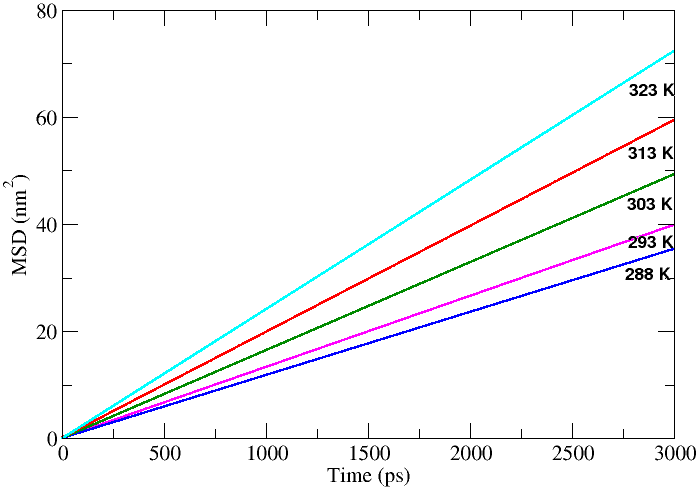}
	\caption{MSD vs time plot of water at different temperatures.}
	\label{msdwatercomb}
\end{figure}

\noindent Figure (\ref{msdwatercomb}) shows the MSD versus time plot at different temperatures. From this figure, it is seen that as the temperature increased slope of the MSD curves have also been increased which in turn increased the self-diffusion coefficient. This is reflected in the following table.

\begin{table}[H]
	\caption{Simulated and experimental values of self-diffusion coefficients of water at different temperatures.}
	\label{table:wat_self_diff}
\resizebox {0.48 \textwidth } {!} {	\begin{tabular}{|c|c|c|c|}
		\hline
		Temperature (K) & \multicolumn{3}{|c|}{Self-Diffusion Coefficient ($10^{-10}$ m$^2$/s)}\\
		\hline
		& Simulated  & Experimental~\cite{manfred}  & Error \\
		& ($D_W^{sim}$) & ($D_W^{exp}$) & (\%)\\
		\hline
		288 & 19.63 $\pm$ 0.01 & 17.66 & 11.16 \\
		\hline
		293 & 22.12 $\pm$ 0.14 & 20.25 & 9.23 \\
		\hline
		303 & 27.38 $\pm$ 0.03 & 25.97 & 5.43\\
		\hline
		313 & 32.95 $\pm$ 0.20 & 32.22 & 2.26\\
		\hline
		323 & 40.19 $\pm$ 0.05  & 39.83 & 0.90\\
		\hline
	\end{tabular}}
\end{table}
\noindent Table (\ref{table:wat_self_diff}) demonstrates that the simulated values of self-diffusion coefficients are in agreement with that of experimental values within 12\% error.

\subsection*{Binary Diffusion Coefficient of Cysteine-water System}

\noindent The self-diffusion coefficient of cysteine and water at particular temperature obtained in previous sections are now used for the calculation of binary diffusion coefficient by means of Darken's relation (\ref{darkeneq}). We have simulated 3 cysteine molecules and 1039 water molecules; 1042 molecules in total. Thus the mole fraction of cysteine is 0.003 and  that of water is 0.997. The values of binary diffusion coefficients obtained after calculation and the experimental value are presented in table (\ref{table:binary_diff}).

\begin{table}[H]
	\caption{Simulated and experimental values of binary diffusion coefficients at different temperatures.}
	\label{table:binary_diff}
\resizebox {0.48 \textwidth } {!} {	\begin{tabular}{|c|c|c|c|}
		\hline
		Temperature (K) & \multicolumn{3}{|c|}{Binary Diffusion Coefficient ($10^{-10}$ m$^2$/s)}\\
		\hline
		& Simulated  & Experimental~\cite{wenrui}  & Error \\
		& ($D_B^{sim}$) & ($D_B^{exp}$) & (\%)\\
		\hline
		288 & 8.20 & 7.90 & 3.80 \\
		\hline
		293 & 9.05 & - & - \\
		\hline
		303 & 10.86 & - & -\\
		\hline
		313 & 13.84 & - & -\\
		\hline
		323 & 16.36 & - & - \\
		\hline
	\end{tabular}}
\end{table}

\noindent From the table (\ref{table:binary_diff}), the comparison of calculated value of binary diffusion coefficient of cysteine in water with that of experimental value available at 288 K in reference~\cite{wenrui} shows an agreement within error of 4\%. To the best of our knowledge, the experimental values of binary diffusion coefficients of cysteine in water at other temperatures are not found till date. However, the calculated values of diffusion coefficients have been increased with increase in temperature. This is because as the temperature increases thermal energy of molecules also increases but density of system decreases; which in turn increases the available space for diffusion. Thus molecular movement in the system become easier. This consequently increases the diffusion coefficient at higher temperatures. By this we can claim that the simulated values of binary diffusion coefficients at higher temperatures are also reasonably good.

\section*{Temperature Dependency of Diffusion}

\noindent As observed in table (\ref{table:binary_diff}) the diffusion phenomenon is strongly dependent on temperature. This temperature dependent behavior of diffusion is given by Arrhenius equation~\cite{udalpaper}:
\begin{equation}
\label{arr equation1}
D = D_0 ~ e^{\frac{\-- E_a}{N_Ak_BT}}.
\end{equation}
In equation (\ref{arr equation1}), $D$ is the diffusion coefficient, $D_0$ represents pre-exponential factor, $E_a$ is the activation energy for diffusion, $N_A$ is Avogardo's number whose value is 6.022 $\times$ 10$^{23}$ mol$^{-1}$, $k_B$ is the Boltzmann's constant whose value is 1.38 $\times$ 10$^{-23}$ JK$^{-1}$~\cite{udalpaper} and $T$ is the absolute temperature.

\noindent On taking natural logarithm in equation (\ref{arr equation1}) we get,
\begin{equation}
\label{arr equation2}
\ln D = \ln D_0 - \frac{E_a}{N_A k_B T}.
\end{equation}

\noindent We obtain the activation energy $E_a$ for diffusion from the slope of $\ln D$ versus ($\frac{1}{T}$) plot (Arrhenius plot) as,

\begin{equation}
E_a = -N_A~k_B \frac{\partial \mathrm{ln~D}}{\partial (1/T)}.
\end{equation}

\noindent The intercept when extrapolated to the 1/$T$ $\rightarrow 0$ in the Arrhenius plot gives the pre-exponential factor.\\

\begin{figure}[H]
	\centering
	\includegraphics[scale=0.305]{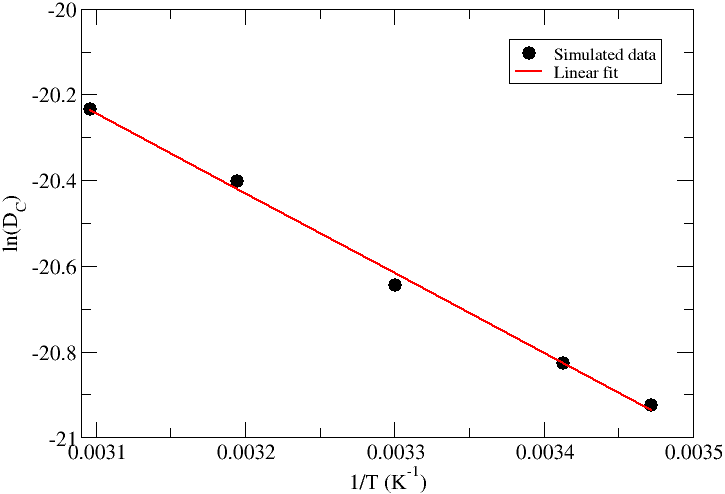}
	\caption{Arrhenius diagram for self-diffusion coefficients of cysteine.}
	\label{arrplotcyssim}
\end{figure}
\noindent Figure (\ref{arrplotcyssim}) depicts the Arrhenius plot of simulated values of self-diffusion of cysteine. The activation energy for self-diffusion of cysteine in water calculated using the slope of linear fit of simulated values is found to be 15.49 kJ mol$^{-1}$.

\begin{figure}[H]
	\centering
	\includegraphics[scale=0.305]{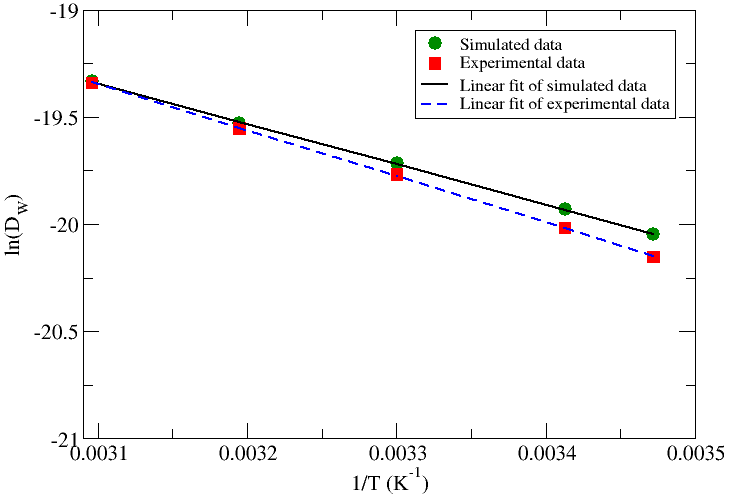}
	\caption{Arrhenius diagram for self-diffusion coefficients of water.}
	\label{arrplotwatsimexpt}
\end{figure}

\noindent Figure (\ref{arrplotwatsimexpt}) portrays the Arrhenius plot of both the simulated and experimental values of self-diffusion of water.The activation energies for self-diffusion of water calculated using the corresponding slope of linear fit of simulated values and experimental values are found to be 15.67 kJ mol$^{-1}$ and 17.88 kJ mol$^{-1}$ respectively. This diagram (\ref{arrplotwatsimexpt}) also shows that the simulated and experimental values of diffusion coefficient are in excellent agreements specially at higher temperatures.

\begin{figure}[H]
	\centering
	\includegraphics[scale=0.305]{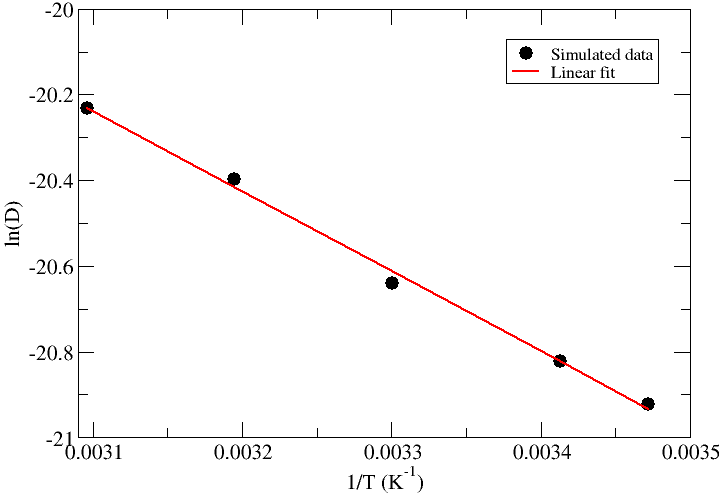}
\caption{Arrhenius diagram for binary diffusion coefficients of binary mixture cysteine and water.}
	\label{arrplotcysbinsim}
\end{figure}

\noindent Figure (\ref{arrplotcysbinsim}) displays the Arrhenius plot of simulated values of binary-diffusion of cysteine in water. The activation energy for binary diffusion of cysteine in water calculated using the slope of linear fit of simulated values is found to be 15.50 kJ mol$^{-1}$.\\

\noindent Figures (\ref{arrplotcyssim}), (\ref{arrplotwatsimexpt}) and (\ref{arrplotcysbinsim}) demonstrate the temperature dependency of diffusion. From these plots it is seen that the diffusion coefficients have increased with temperature. We have calculated the activation energies for diffusion of cysteine, water and their binary mixture by using the slopes of respective Arrhenius plots which are tabulated below.

\begin{table}[H]
	\centering
	\caption{Activation Energies for Diffusions.}
	\label{table:activationenergy}
	\begin{tabular}{|c|c|c|c|}
		\hline
		System  & \multicolumn{3}{|c|}  {Activation Energy ($E_a$) (kJ mol$^{-1}$)} \\
		\hline
		& Simulated & Experimental& Error (\%)\\
		\hline        
		Water  & 15.67 & 17.88~\cite{manfred} & 12.36 \\
		\hline
		Cysteine & 15.49  & &  \\
		\hline
		Binary mixture  & 15.50  & & \\
		\hline
	\end{tabular}
\end{table}
\noindent From this table (\ref{table:activationenergy}), it is observed that the activation energies for self-diffusion of cysteine and for the binary diffusion of cysteine in water are almost same. This implies that the concentration of cysteine in the system is infinitesimal. Further the activation energy calculated for simulated and experimental values of self-diffusion of water are in agreement within the error of 13\%.

\section*{Conclusions and Concluding Remarks}
\noindent In the present work, we performed the molecular dynamics study of transport property- diffusion of 3 cysteine molecules in 1039 SPC/E water molecules at 288 K, 293 K, 303 K, 313 K and 323 K temperatures using GROMACS 4.6.5 package. We used OPLS-AA force field parameters throughout the simulation. The system under study was confined in a cubic box having one side 3.2 nm after making topology file. To account the short-range interactions cut-off distance of 1.0 nm was used where as PME was employed for long-range interactions. All bonds were kept fixed by LINCS algorithm while the bond angles were free to vibrate.

\noindent The energy minimization was carried out using steepest descent method with 0.01 nm step size. After energy minimization, we proceeded towards equilibration run (NPT ensemble), where temperature and pressure couplings were done by using velocity rescaling and Berendsen barostat respectively. The equilibration run  was performed for 50 ns taking 0.001 nm step size; which brought the system in equilibrium.
The stable system obtained after equilibration run was subjected for production run (NVT ensemble). It was also done for 50 ns and 0.001 nm step size. The production run yielded the important results which were analyzed through XMGRACE plots. Various energies of the system were portrayed in the energy profile diagrams. The structures of solute and solvent of the system were studied via radial distribution functions. The analysis of RDF plots at different temperatures reveal that the system becomes less organized with increase in temperature. 

\noindent The transport property of system, diffusion, was studied. The self-diffusions of both cysteine and water were separately determined by using the Einstein's equation. The diffusion of binary mixture of cysteine and water was calculated by means of Darken's relation. The simulated values obtained were then compared with corresponding experimental values. The simulated values of self-diffusion coefficients of water were showed excellent agreements with that of experimental values specially at higher temperatures and small deviation ($\sim$11\%) at low temperature (288 K). Likewise, the simulated values of binary diffusion coefficient of cysteine in water was compared with that of available experimental value at 288 K. This comparison showed very little deviation of about 4\%. Further the values of diffusion coefficients at other temperatures showed correlation with temperature i.e. higher at higher temperatures. Moreover, the diffusion coefficients of cysteine were smaller than that of water for each corresponding temperature. This is because cysteine molecule is much larger than water molecule, so cysteine suffers more hindrance than water while moving in the system. Thus we can conclude that the simulated values are reasonably good. The temperature dependency of diffusion were studied through Arrhenius plots. The Arrhenius plots were also utilized to calculate the activation energies for diffusion. The simulated and experimental values of activation energies for self-diffusion of water were in agreement within 13\% error.

\noindent We can inferred that the molecular dynamics simulation is very reliable tool for the study of dynamic properties of a system. It is convenient from both time and money and it is free from experimental hazards. Also this study of transport property- diffusion, of cysteine in water will be reference to further study. In near future, we are interested to extend the present study to diffusion of cysteine in heavy water, with varying concentration as well as temperature, and compare the corresponding values in normal water. We are also curious to study the other transport properties such as viscosity and thermal conductivity in the present system.

\section*{Acknowledgements}
\noindent HB acknowledges the financial support from University Grants Commission (UGC), Nepal for M.Sc. thesis. NA acknowledges CRG support from 
UGC award no. CRG 2073/74 S\&T -01. We also acknowledge the support from ICQ-13 for Computational facilities.

\end{document}